\documentclass[12pt]{article}

\newenvironment{sciabstract}{%
\begin{quote} \bf}
{\end{quote}}

\newcounter{lastnote}

\newcommand{\urom}[1]{\uppercase\expandafter{\romannumeral#1}}

\usepackage{physics}
\usepackage{graphicx} 
\usepackage{amsmath}

\title{Electronic transport close to boundaries of semi-infinite graphene and their interfaces}

\author
{Fanbing Xia,$^{1\ast}$ Jian Wang $^1$ \\
\\
\normalsize{$^{1}$Department of Physics, The University of Hong Kong}\\
\normalsize{$^\ast$E-mail:  c20040325@gmail.com}\\
}

\date{}

\begin{document} 


\baselineskip24pt


\maketitle


\begin{sciabstract}
  Abstract: Transport properties of 2D materials especially close to their boundary has received much attention after the successful fabrication of graphene and other fascinating materials afterwards. While most previous work is devoted to the conventional lead-device-lead setup with a finite size center area, this project investigates real space transport properties of infinite and semi-infinite 2D system under the framework of Non-equilibrium Green’s function. The commonly used method of calculating the Green’s function by inverting a matrix in the real space directly can be unstable in dealing with large systems as sometimes it gives non-converging result. Not to mention that the calculation error and time increase drastically with size of the system. By transforming from the real space to momentum space, the author managed to replace the matrix inverting process by Brillouin Zone integral process which can be greatly simplified by the application of contour integral. Combining this methodology with Dyson equations, we are able to calculate transport properties of semi-infinite graphene close to its zigzag boundary and its combination with other material including s-wave superconductor. Interference pattern of transmitted and reflected electrons, graphene lensing effects and difference between Specular Andreev reflection and normal Andreev reflection are verified through our calculation. 
\end{sciabstract}


\section{Introduction}
As the emergence of a variety family of 2-D materials in the past decades, their transport properties has become one of the central focuses of condensed matter physicists. Graphene in particular has received much attention because of its Dirac like low energy excitations, gapless dispersion relation and degenerated boundary states. For infinitely large graphene system, different approaches has been used in calculating the conductance of graphene in the existence of disorder[1-4]. There are also plenty of research papers devoted to the investigation of transport properties of graphene nanoribbons with zigzag or armchair boundary[5,6]. These research more frequently utilize the finite size of the system in at least one direction to apply scattering matrix approach or Non-equilibrium Green's function. In this paper, the authors use NEGF to investigate point to point transport properties of infinite, semi-infinite graphene and their coupling with 2-D superconductors. Unlike those focused on the low energy Dirac like excitation in graphene, our method is applicable to the full range of dispersion which indicates that this method is applicable to most 2-D material as long as we have the tight-binding description of the system.

\section{Green's function of infinitely large pristine graphene}
The calculation method[7] and physical model[8,9] we will introduce in this section is similar to what proposed by two previous papers. The physical system of interest contains three parts, two STM probes and a piece of infinitely large graphene. The main quantity we are calculating is the real space retarded Green's function $G^R(\textbf{r}_1,\textbf{r}_2,E)$ because it is the main ingredient we need to calculate electron transmission coefficient. The parameter $E$ is often omitted and Green's function will mean retarded Green's function if there is no specification throughout the paper.

The first order tight-binding Hamiltonian of graphene written in its momentum space is 
\begin{equation}
\hat{H} = \sum_{\textbf{k}} \epsilon_{\textbf{k}}^+ c^\dagger_{\textbf{k}} c_{\textbf{k}}+\epsilon_{\textbf{k}}^- c^\dagger_{\textbf{k}} c_{\textbf{k}}
\end{equation}
where $\epsilon_{\textbf{k}}^+ = \sqrt{1+4cos^2(\frac{a}{2}k_y)+4cos(\frac{\sqrt{3}a}{2}k_x)cos(\frac{a}{2}k_y)}$ and $\epsilon^-_{\textbf{k}} = -\epsilon^+_{\textbf{k}}$. This can be easily obtained from diagonalizing the 2 by 2 off diagonal Hamiltonian
\begin{equation}
\hat{H} = \sum_{\textbf{k}}
\begin{bmatrix}
\
0 & f(\textbf{k})\\
f^*(\textbf{k}) & 0
\end{bmatrix}
\end{equation}
where $ f(\textbf{k}) = 1 + e^{i\textbf{k}\textbf{a}_1} +e^{i\textbf{k}\textbf{a}_2}$. From the diagonalized Hamiltonian, the Green's function of pristine graphene can be easily achieved
\begin{equation}
\hat{G}^R = \sum_{\textbf{k}}
\begin{bmatrix}
\
\frac{1}{E+i0_{+}-\epsilon^+_{\textbf{k}}} & 0\\
0 & \frac{1}{E+i0_{+}-\epsilon^-_{\textbf{k}}}
\end{bmatrix}
\end{equation}
In order to calculate the real space Green's function, we have to switch back from the diagonal basis back to its sublattice basis
\begin{equation}
G(\textbf{r}_2,\textbf{r}_1) = \sum_{\textbf{k}} \frac{e^{i\textbf{k}(\textbf{r}_1-\textbf{r}_2)}}{E^2-\epsilon_{\textbf{k}}^2 + i0_E}
\begin{bmatrix}
\
E & tf(\textbf{k}) \\
tf^*(\textbf{k}) & E
\end{bmatrix}
\end{equation}
and since we are now talking about the infinitely large graphene with full translational symmetry, the summation over $\textbf{k}$ can be transformed naturally to the integral over first Brillouin zone. The integral is calculated from contour integral. Conventionally the shape of the first Brillouin zone of 2-D honey-cone structure is hexagonal as shown in the left figure of fig.[1]. However, since the k-space is invariant under translation of reciprocal lattice vectors, we are able to reform the hexagon into a rectangle, as indicated in fig.[1] which can greatly simplify our calculation. 
\begin{figure}[htb]
    \includegraphics[scale = 0.25]{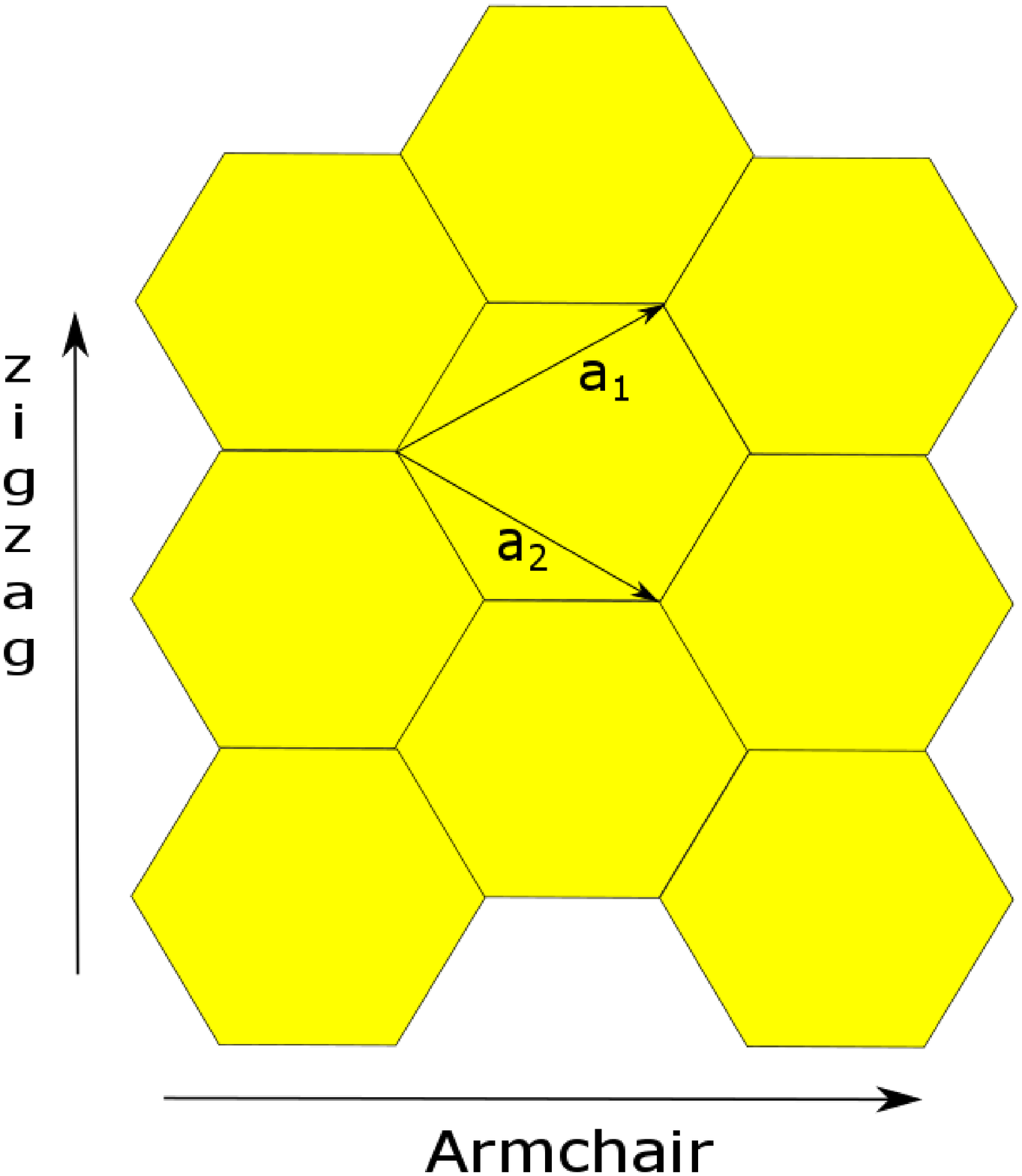}
    \includegraphics[scale = 0.25]{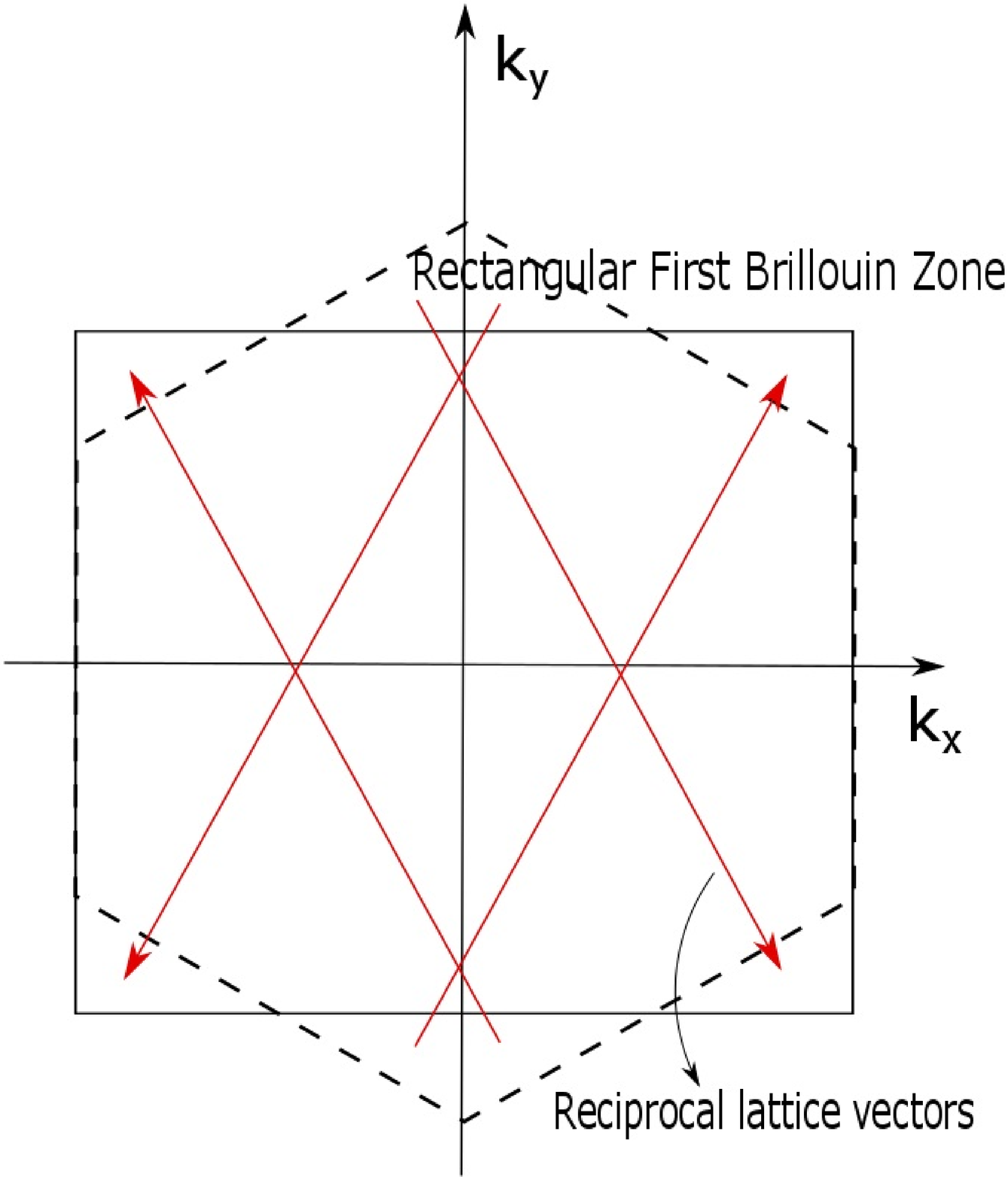}
    \caption{The left figure shows the lattice structure of graphene and the right shows how the Brillouin zone is transformed into a rectangular shape.}
    \label{fig:mesh1}
\end{figure}

Because of the symmetry of the dispersion relation, let's discuss the case for $0<E<t$ first. The case $|E|>t$ will not be discussed since Fermi surface of pristine Graphene is close to its Dirac point which means it would be rather different to move the Fermi-level very far from the Dirac point. Now that the range of $k_y$ and $k_x$ are independent, we can integrate out the $k_x$ degree of freedom first which can be done easily by contour integral. The way of drawing the contour is as described in fig.[2]. We extend the integral path to complex plane from $\pm\frac{2}{\sqrt{3} a}\pi$ to infinity $\pm\frac{2}{\sqrt{3} a}\pi +\infty i$ and thus constructed a rectangular contour. The integral of the right and left branches cancel out each other, and the integral at the infinity is zero. As a result we can apply Residue theorem to calculate the integral over $k_x$ direction. Although not mentioned in the original text [7], this step also involves some subtleties. This is because the poles actually fall symmetrically on the real axis and if we take into account both of them, we will lose the real part of retarded Green's function. The asymmetric way of picking poles arises from the infinite small imaginary part in the denominator of eqn.(4) as it excludes one side of the poles out of the contour, as indicated in the middle graph of fig.[2].

\begin{figure}[htb]
	\centering
    \includegraphics[scale = 0.18]{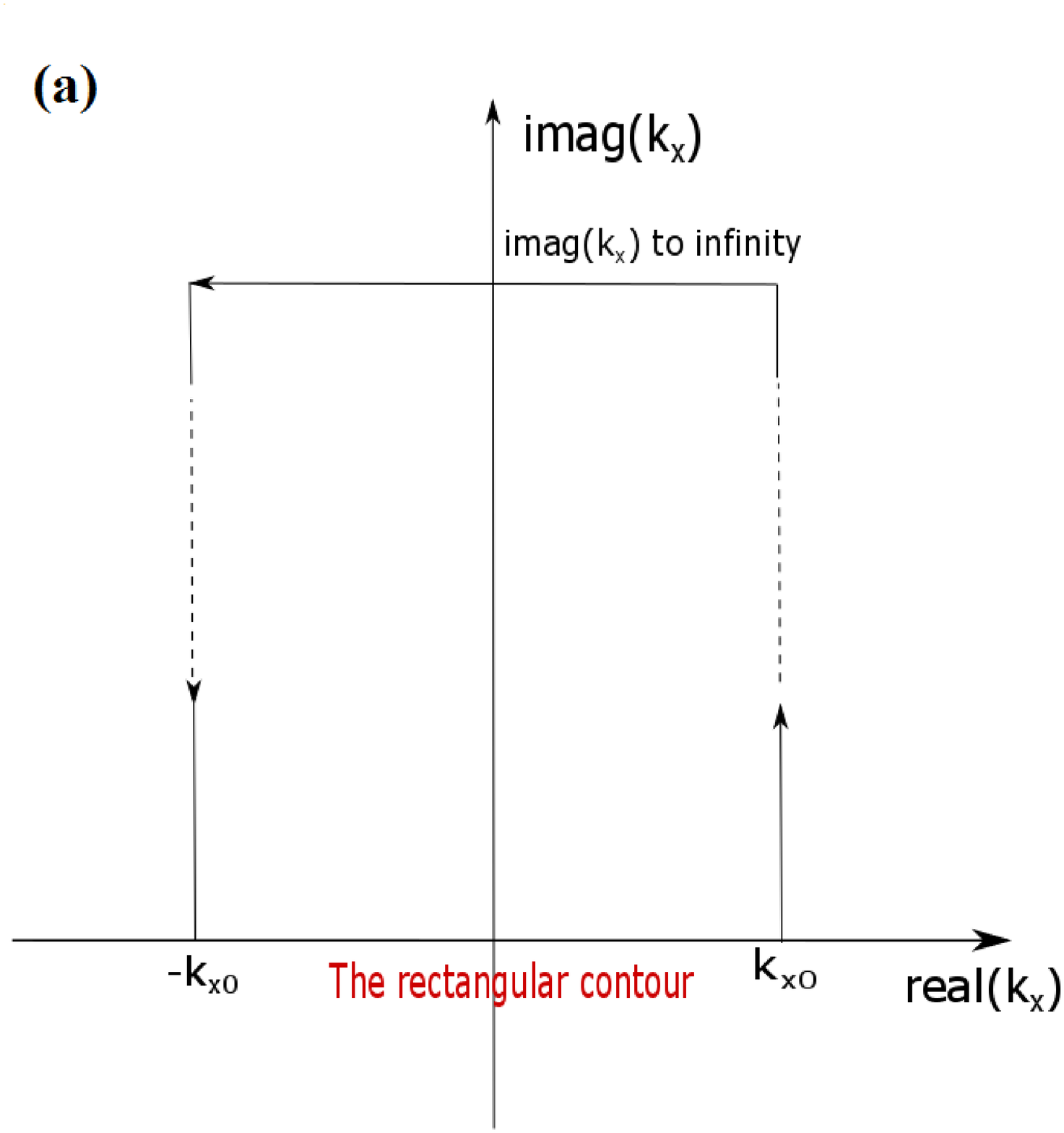}
    \includegraphics[scale = 0.18]{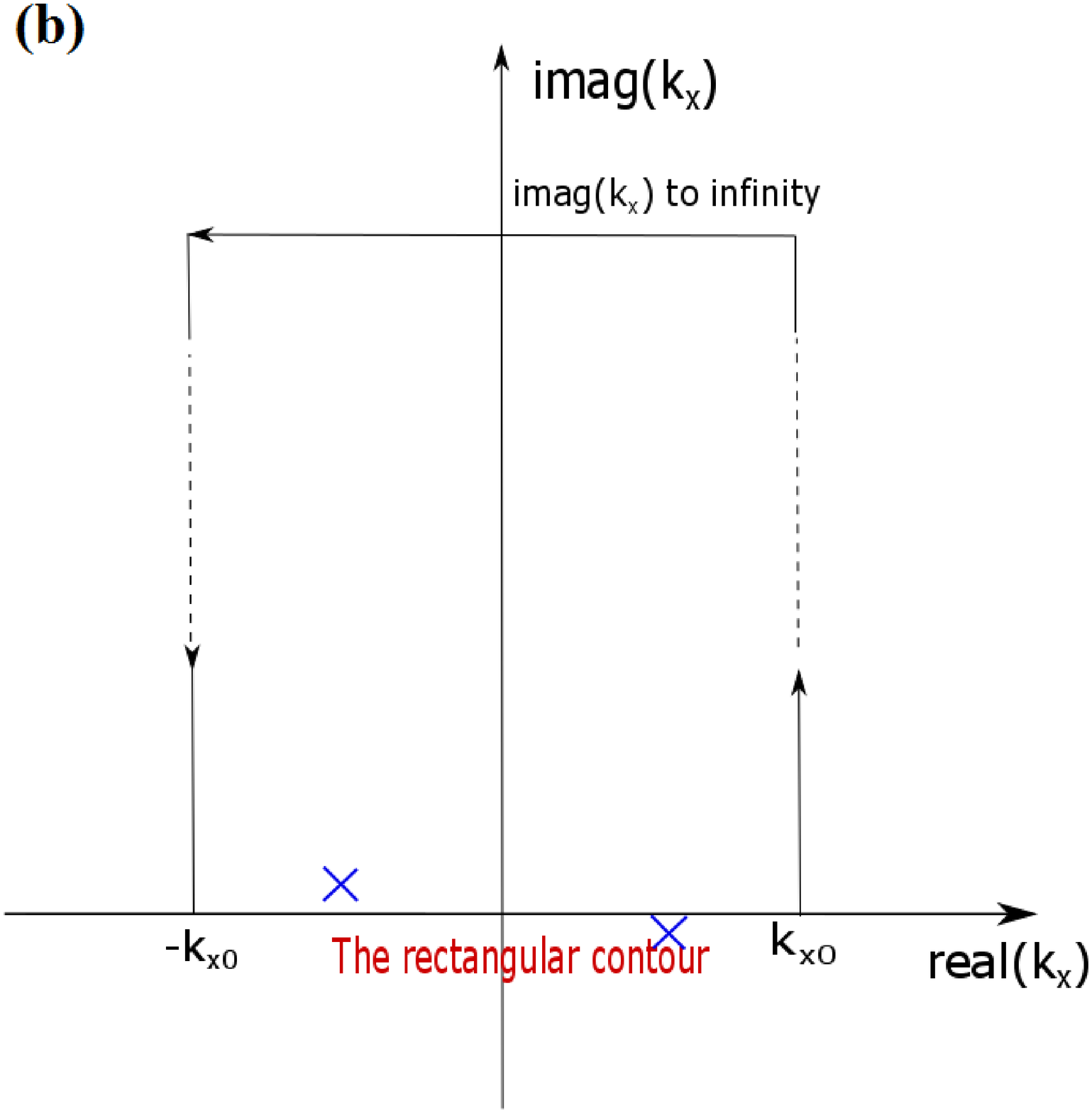}
    \includegraphics[scale = 0.18]{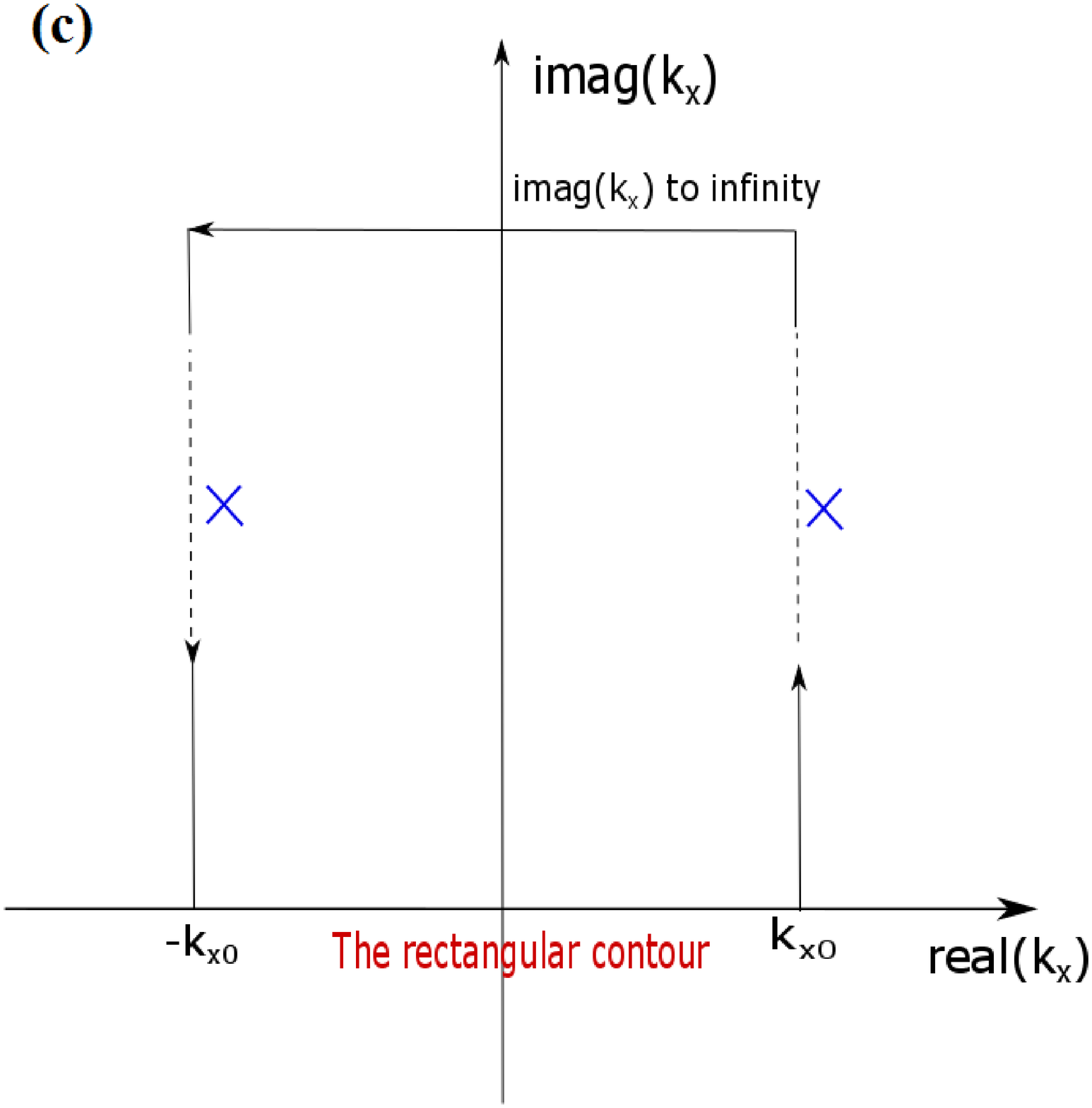}
    \caption{The integral in one direction is simplified by Residue theorem with a contour indicated as above. The crosses in (a) and (b) represent the position of the poles after taking into account the effect of the infinitesimal part in the denominator of retarded Green's functions. Notice that even if there is no pole on the real axis, we are still able to calculate the integral with exact result by finding poles in the imaginary plane as indicated in (c). }
    \label{fig:mesh1}
\end{figure} 

As a result, the simplified Green's function can be written as
\begin{equation}
G(\textbf{r}_2,\textbf{r}_1) = \frac{a}{2\pi}\int_{-\pi}^{\pi} dk_y \frac{ie^{i\textbf{k}(\textbf{r}_1-\textbf{r}_2)}}{4\sin(\frac{\sqrt{3}a}{2}k_x)\cos(\frac{k_y}{2})}
\begin{bmatrix}
\
E & tf(\textbf{k}) \\
tf^*(\textbf{k}) & E
\end{bmatrix}
\end{equation}
where 
\begin{equation}
k_x = 
\begin{cases}
-\frac{2}{a\sqrt{3}}|\acos(\frac{E/t-1-4\cos^2(\frac{ak_y}{2})}{4\cos(\frac{ak_y}{2})})|,    & 0<E<t \\
\frac{2}{a\sqrt{3}}|\acos(\frac{E/t-1-4\cos^2(\frac{ak_y}{2})}{4\cos(\frac{ak_y}{2})})|,      & -t<E<0
\end{cases}
\end{equation}
The left integral in the $k_y$ direction can then be calculated using numerical methods because its poles are of order less than 1 which does not survive under integral. Moreover, if we are interested in Green's function between two points far apart from each other, we can use stationary phase approximation to further simplify our calculation[6].

\section{Green's function of semi-infinite Graphene close to its zigzag boundary}

The Green's function of semi-infinite graphene with an armchair boundary has been calculated using the symmetry property of the wavefunction [10]. Here we use a different approach to calculate the Green's function close to the zigzag boundary, and as we will see soon, its coupling with other materials. The building block we will be using to calculate these Green's functions is the surface Green's functions. To calculate this quantity, we need to make use of the translational symmetry in $y$ direction and Dyson equations. As indicated in fig.[3], the combination of two semi-infinite graphene sheet must recover all the physical quantities of an infinitely large graphene. Therefore, we must have the following Dyson equation

\begin{equation}
\begin{split}
\bra{x_0,y_1}\hat{G}\ket{x_0,y_2} = &\bra{x_0,y_1}\hat{g}_s\ket{x_0,y_2} + \\&\sum_{y',y''} \bra{x_0,y_1}\hat{g}_s\ket{x_0,y'}\bra{x_0,y'} \hat{T}
\ket{x_0',y'}\bra{x_0+a/\sqrt{3},y'} \hat{g}_s\ket{x_0',y''} \\ &\bra{x_0',y''} \hat{T}^\dagger\ket{x_0,y''}\bra{x_0,y''} \hat{G}\ket{x_0,y_2}
\end{split}
\end{equation}
where $x_0$ stands for the boundary position of the left graphene and $x_0'$ stands for the that of the right graphene. Fourier transform of this equation gives
\begin{equation}
G_{BB}(x_0,x_0,k_y) = g_{s,BB}(x_0,x_0,k_y) + t^2g_{s,BB}(x_0,x_0,ky)^2G(x_0,x_0,ky).
\end{equation}
In eqn.(7) all the Green's functions are still 2 by 2 matrices (A,B sublattice) while in eqn.(8) we are focusing on a particular element of the matrix. Eqn.(8) is a quadratic equation of $g_{s,BB}(x_0,x_0,k_y)$ which has two solutions. To determine which solution to choose, we developed a second method to solve for $g_{s,BB}(x_0,x_0,k_y)$ and choose the common solution of these two methods. This second method is very similar to the first one. We attach an infinitely long atomic chain to the zigzag boundary of the semi-infinite graphene and using the fact that the combined system must recover all physical quantities of a semi-infinite Graphene, we can write down another Dyson equation that has one common solution with the previous one. Since the second method is very similar to the first one, we here will provide the final result
\begin{equation}
g_{s,BB}(x_0,x_0,ky) = 
\begin{cases}
\begin{split}
&1+2\cos(\frac{ak_y}{2})\cos(\frac{\sqrt{3}ak_x}{2})\\&-i2\sin(\frac{\sqrt{3}ak_x}{2})\cos(\frac{ak_y}{2}),
\end{split}      &       0<E<t \\
\begin{split}
&-1-2\cos(\frac{ak_y}{2})\cos(\frac{\sqrt{3}ak_x}{2})\\&-i2\sin(\frac{\sqrt{3}ak_x}{2})\cos(\frac{ak_y}{2}),      
\end{split}&             -t<E<0
\end{cases}
\end{equation}
where $k_x$ maintains the relation eqn.(6) with ky.
\begin{figure}[htb]
    \includegraphics[scale = 0.4]{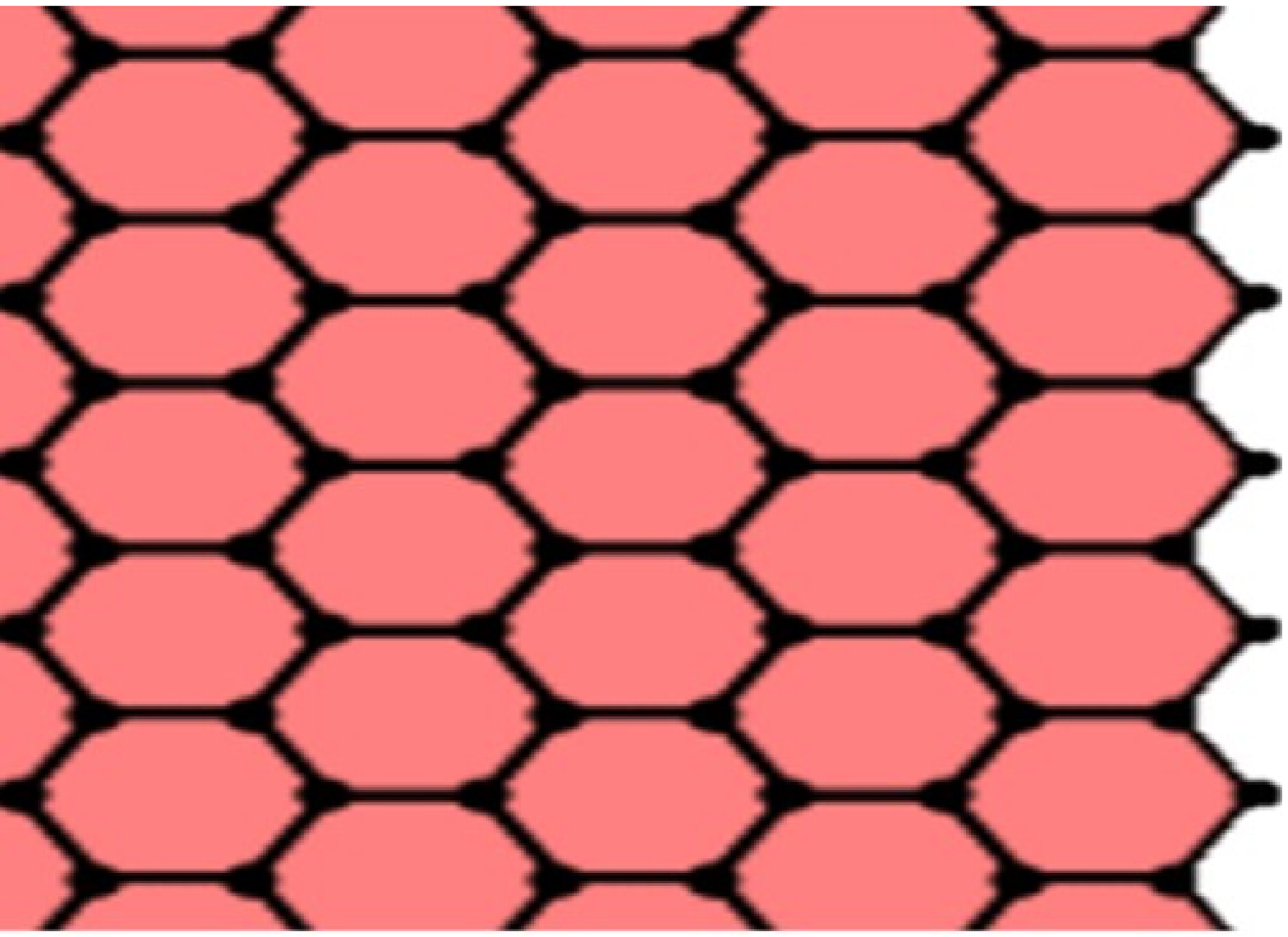}
    \includegraphics[scale = 0.33]{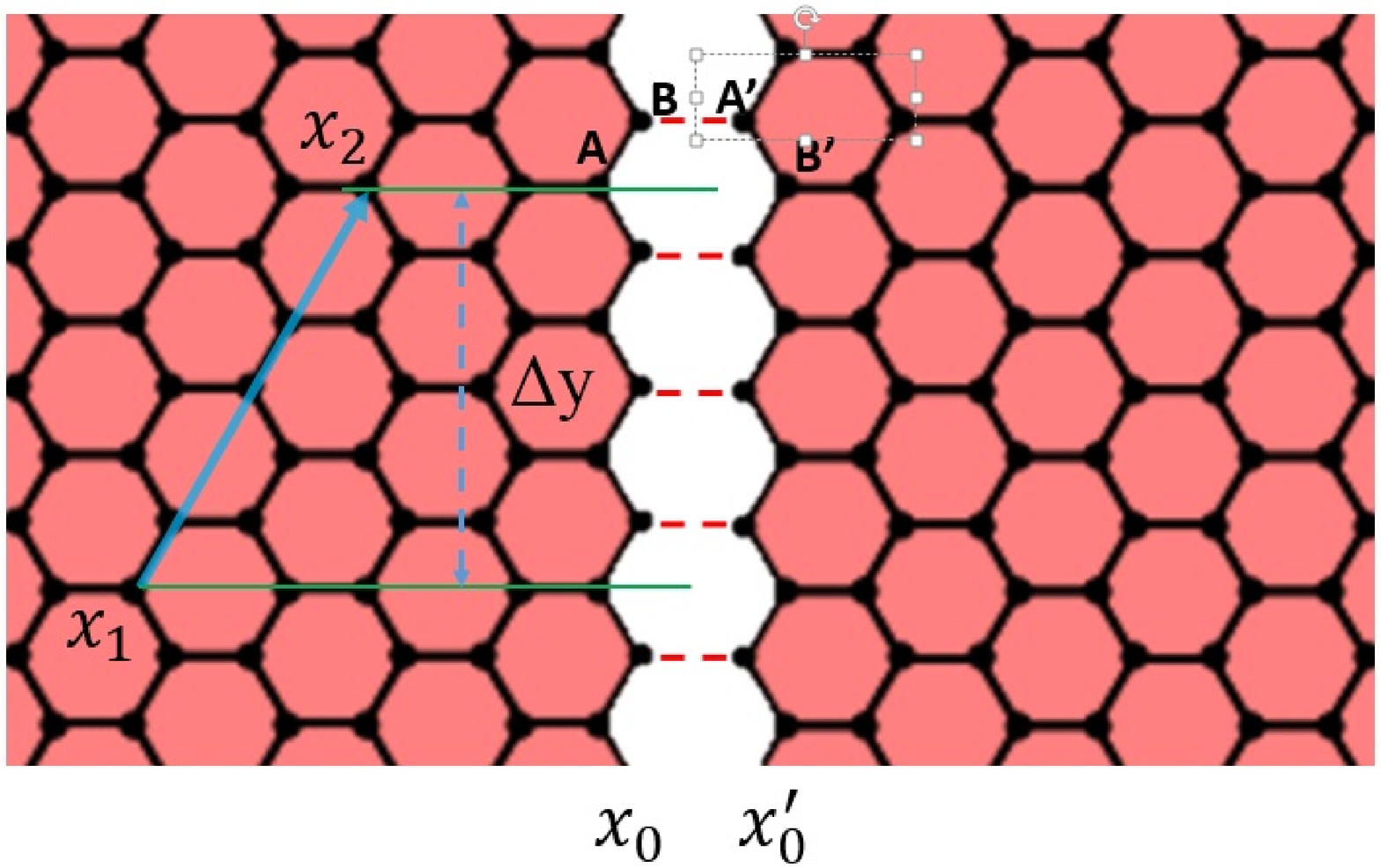}
    \caption{The left figure shows the geometry of a semi-infinite graphene at its zigzag boundary. The right figure shows how two semi-infinite graphene must recover an infinitely large graphene. The right-most atom of the left semi-infinite graphene is the position where we defined as $x_0$}
    \label{fig:mesh1}
\end{figure}

With the surface Green's function in hand, we are ready to calculate other physically interesting quantities. Firstly we are interested in the Green's function away from the boundary
\begin{equation}
\begin{split}
G_{BB}(x_1,x_2,ky) = &g_{s,BB}(x_1,x_2,ky) +\\&t^2g_{s,BB}(x_1,x_0,ky)g_{s,BB}(x_0,x_0,ky)G_{BB}(x_0,x_2,ky)
\end{split}
\end{equation}
\begin{equation}
\begin{split}
G_{BB}(x_1,x_0,ky) = &g_{s,BB}(x_1,x_0,ky) +\\&t^2g_{s,BB}(x_1,x_0,ky)g_{s,BB}(x_0,x_0,ky)G_{BB}(x_0,x_1,ky)
\end{split}
\end{equation}
The above two equations are sufficient to solve for $g_{s,BB}(x_1,x_0,ky)$ and $g_{s,BB}(x_1,x_2,ky)$ and other matrix elements can also be calculated straight forwardly. The real space Green's function therefore writes 
\begin{equation}
g_s(x_1,x_2,\bigtriangleup y) = \frac{a}{2\pi}\int^{\pi}_{-\pi} dk_y g_{s}(x_1,x_2,ky) e^{iky\bigtriangleup y}
\end{equation}
which can be calculated easily from numerical methods.

\section{Coupling between two semi-infinite systems}

In this section we use the method developed in last section to investigate two different systems. The first one is coupling between two semi-infinite graphene with different Fermi levels. It has been experimentally verified that if we set Fermi level of the left Graphene higher than the Dirac point and that of the right Graphene lower (a PN junction), the electron transmitting through the boundary from the left plane will focus to one point in the right plane, namely the so called graphene lensing effect[11]. While previous simulations either utilize the finite size of a graphene ribbon [12] or the approximation that the coupling will not greatly affect the graphene dispersion relation significantly, we are able to give a result without applying the above approximations. The Dyson equation for $G(x_l,x_r,k_y)$

\begin{equation}
G(x_l,x_r,ky) = g_{s,E_l}(x_l,x_0,ky)\hat{T}G(x_0',x_r,ky)
\end{equation} 
where $G(x_0',x_r,ky)$ can be written as
\begin{equation}
\begin{split}
G(x_0',x_r,ky) =&g_{s,E_r}(x_0',x_r,ky)+ \\&g_{s,E_r}\hat{T}(x_0',x_0',ky)g_{s,E_l}\hat{T}^{\dagger}(x_0,x_0,ky)G(x_0',x_r,ky)
\end{split}.
\end{equation}

The Green's function in eqn.(13) and eqn.(14) are 2 by 2 matrices with respect to its sublattices and the transmission matrix writes
\begin{equation}
\hat{T} = 
\begin{bmatrix}
\
0 & 0\\
1 & 0
\end{bmatrix}.
\end{equation}
Here $G$ stands for the Green's function of the combined system and $g_{s,E_l/E_r}$ stands for the semi-infinite Green's function of the left/right graphene. In fig.[4] we show the graphene lensing effect from our calculation. The focusing strength and the focal length depends solely on the relative Fermi levels. The data we show here demonstrates the case which the left Fermi surface is fixed at $0.5t$ above the Dirac point. When the right Fermi surface is set $0.5t$ below the Dirac point, we have a maximized  focusing strength and the focal length is equal to the distance between the source probe and the boundary. When the right Fermi surface moves closer to/farther from the Dirac point, the focal length will become longer/shorter and the focusing effect will be weaken drastically. This is consistent with the semi-classical arguments. 

\begin{figure}[htb]
	\centering
    \includegraphics[scale = 0.18]{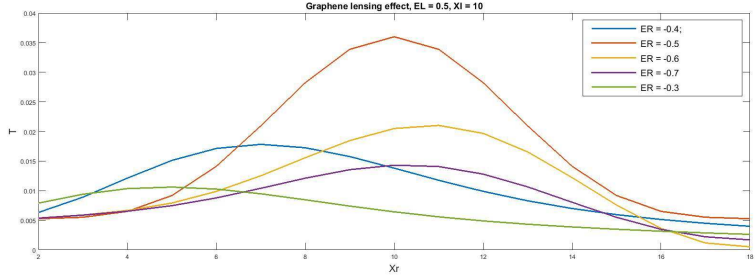}
    \caption{This plot shows the transmission between STM probes on different sides of Graphene-Graphene structure. The probe on the left is fixed at a position with distance $10\sqrt{3}a$ away from the boundary and the other probe is moving around on the right plane. The Fermi surface of the left graphene is set 0.5t above its Dirac point and the Fermi surface of the right graphene is chosen to be five values below the Dirac point of the right graphene.}
    \label{fig:mesh1}
\end{figure} 

Another system we think that is interesting enough to investigate is the coupling between graphene and conventional s-wave superconductor because of the proposed specular Andreev reflection[13]. Firstly we will need to write down that BdG hamiltonian of graphene and the superconductor. For simplicity, we assume the superconductor to have a square lattice structure and therefore its BdG hamiltonian writes
\begin{equation}
H_{BdG} = \Sigma_{\textbf{k}}
\begin{bmatrix}
\epsilon_{\textbf{k}} -E_F & \bigtriangleup \\
\bigtriangleup^* & E_F-\epsilon_{-\textbf{k}}
\end{bmatrix}
\end{equation}
where $\epsilon_{\textbf{k}} = U-2t[\cos(ak_x)+\cos(ak_y)]$ is the energy dispersion of the square lattice and $\bigtriangleup$ is the superconductor energy gap. To find the Green's function of the 2D superconductor we only need to solve the equation
\begin{equation}
\hat{G}^R_{Sc} [E+i0_+-\hat{H}] = I
\end{equation}
which gives us the explicit expression of the real-space Green's function in its Nambu representation
\begin{equation}
G^R_{Sc}(\textbf{r}_1,\textbf{r}_2) =
\sum_{\textbf{k}} e^{i\textbf{k}(\textbf{r}_2-\textbf{r}_1)}
\begin{bmatrix}
\
\frac{E+\epsilon_{\textbf{k}}-E_F}{E^2-(\epsilon_{\textbf{k}}-E_F)^2 -\bigtriangleup^2+ i0_E} & \frac{\bigtriangleup}{E^2-(\epsilon_{\textbf{k}}-E_F)^2-\bigtriangleup^2 + i0_E} \\
\frac{\bigtriangleup^*}{E^2-(\epsilon_{\textbf{k}}-E_F)^2-\bigtriangleup^2 + i0_E} &\frac{E-\epsilon_{\textbf{k}}+E_F}{E^2-(\epsilon_{\textbf{k}}-E_F)^2 -\bigtriangleup^2+ i0_E}
\end{bmatrix}.
\end{equation}
The Green's function of Pristine graphene becomes a 4 by 4 matrix under Nambu representation and so does the Green's function of semi-infinite graphene. For normal graphene the superconducting energy gap is zero and therefore the extra Hamiltonian and Green's function for the hole degree of freedom is simply a copy of that of electrons' with inverted energy and $\textbf{k}$. Now as indicated in fig.[3], we assume that only the B sublattice of Graphene is coupled to the superconductor so that we are not interested in the A sublattice along the boundary of Graphene which make it reasonable for us to again focus on the $BB$  or $AB$ element of Graphene Green's function. Under Nambu representation each element is still a 2 by 2 matrix with the hole degree of freedom included. 

The surface Green's function of the semi-infinite superconductor is calculate by the method introduced in previous section. In order to observe the transmission signature of specular Andreev reflection, we put both the source and drain STM probe on the graphene side and calculate the Andreev transmission coefficient using the relation 
\begin{equation}
T_A(E) = Tr[\Gamma_{1,ee}(E) G^R_{12,eh}(E) \Gamma_{1,hh}(-E) G^A_{21,he}(E)]
\end{equation}
where $\Gamma_1$ and $\Gamma_2$ are the bandwidth functions of two probes [9,15]. Notice that the trace here is actually unnecessary as each probe is only coupling the graphene sheet with one single atom. The result of our calculation is given in fig.[5]. We have fixed the position of the source probe and the superconducting energy gap while changing the Fermi level and energy of the incident electron. As was theoretically predicted, with in the energy gap of the superconductor, when the reflected hole falls in the valence band of graphene, the specular Andreev reflection will happen. In (a) of fig.[6], $E_F$ is set at the Dirac point and the energy of the electron is set smaller than the energy gap, we observe that Andreev transmission coefficient peaks around the source probe. In figure (b) and (d) the energy of the electron is set larger than the energy gap and therefore a much weaker Andreev reflection is observed. In (c) of fig.[5], even though the electron energy falls within the energy gap, the reflected hole is still in the conduction band as we have raised the Fermi level. In such a case normal Andreev reflection is expected to happen. 

\begin{figure}[htb]
	\centering
    \includegraphics[scale = 0.28]{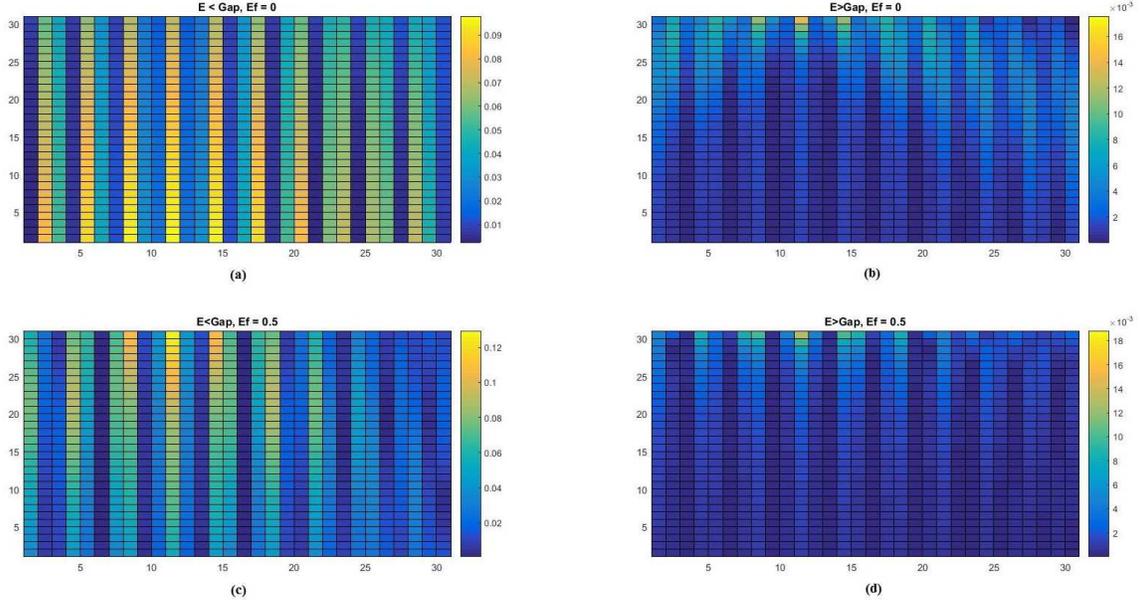}
    \caption{Andreev transmission map of a graphene coupled to a superconductor. The Boundary is a the top and the source probe is fixed at the grid(10,20). Specular Andreev reflection is expected to happen in (a) and normal Andreev reflection is expected to happen in (c).}
    \label{fig:mesh1}
\end{figure} 

\section{Summary and discussion}
	So far we have introduced an approach of calculating the real-space Green's functions of semi-infinite 2D system from that of the translational invariant infinite system. Moreover, with the help of Dyson equations we are able to calculation Green's function of two different 2D systems that are glued together. 
 
\begin{figure}[htb]
	\centering
    \includegraphics[scale = 0.28]{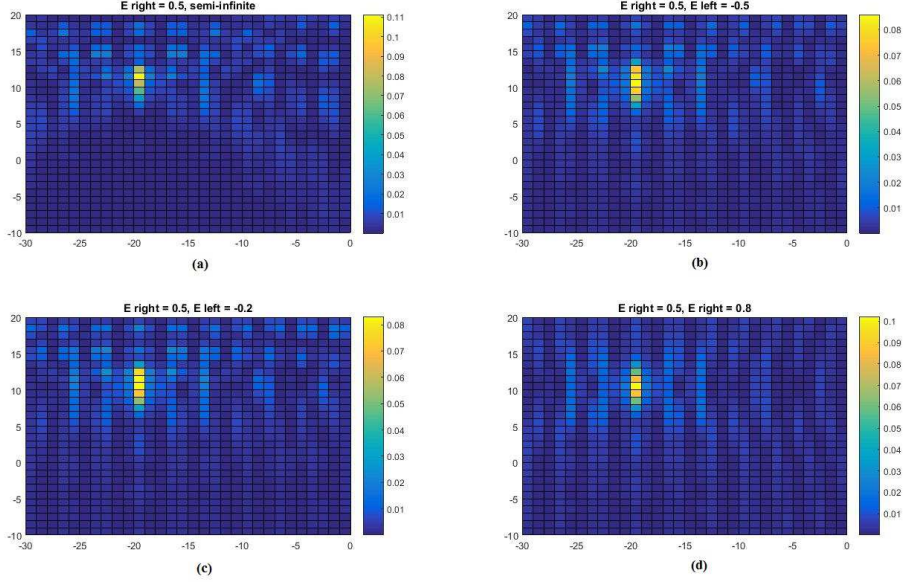}
    \caption{Electron transmission map of a graphene PN structure with the source probe fixed at (10,20) and the drain probe moving around.}
    \label{fig:mesh1}
\end{figure}

	Then we use the method to investigate two physical systems that were previously predicted to behave interestingly from semi-classical and wave function analysis. From our calculation we found an obvious focusing effect for electrons passing through a graphene PN junction, and the relation between focal length and Fermi levels is consistent with semi-classical prediction. For the case of graphene coupling to a superconductor we observe that firstly, Andreev transmission solely come from the reflection at the boundary because we cannot observe any unique transmission around the source probe as in normal transmission (in fig.[6] the transmission coefficient around the source probe is clearly larger). Secondly, when the energy of the electron is larger than the energy gap, electrons are much more likely to transmit through the boundary as a single particle excitation and therefore the Andreev transmission drop drastically. For electron with energy below the gap, if the reflected hole falls in the conduction band, the transmission coefficient will drop as the drain probe moves away from the boundary or moves away from the source probe horizontally. If the reflected hole falls in the valence band, due to specular Andreev reflection, the transmission coefficient maximizes away from the boundary probably as an result of the interference of reflected holes. 
	
	The method introduced in second and third sections on computing Green's functions apply not specifically to graphene but to any 2D material with a known band structure. In fact, we have applied this method to 1D system as well in calculating the surface Green's function of the STM probes as well.

\section{Reference}
1. N. M. R. Peres, Reviews of Modern Physics, 82(3), 2673 (2010).

2. E. H. Hwang, and S. D. Sarma. Physical Review B 77, 19 (2008).

3. Shon, Nguyen Hong, and Tsuneya Ando. Journal of the Physical Society of Japan 67, 7 2421-2429 (1998). 

4. Ando, Tsuneya. Journal of the Physical Society of Japan 75, 7: 074716.  (2006).

5. Wakabayashi, Katsunori, Yositake Takane, Masayuki Yamamoto, and Manfred Sigrist. "Electronic transport properties of graphene nanoribbons." New Journal of Physics 11, 9, 095016 (2009).

6. X.Du, S. Ivan, B. Anthony, and Y. A. Eva . Nature nanotechnology 3, no.8 (2008)	.

7. S. R. Power, and M. S. Ferreira. Physical Review B 83, 15, 155432 (2011).

8. M. Settnes, R. P. Stephen  P. Dirch , and A.P. Jauho, Physical Review B 90, 3, 035440 (2014).

9. M. Settnes, R. P. Stephen  P. Dirch , and A.P. Jauho, Physical review letters 112, 9, 096801 (2014).

10. J. M. Duffy, P. D. Gorman, S. R. Power, and M. S. Ferreira. Journal of Physics: Condensed Matter 26, 5, 055007 (2013).

11. V. V. Cheianov, F. Vladimir, and B. L. Altshuler. Science 315, 5816 1252-1255 (2007). 

12. Y. Xing, J. Wang, and Q. F. Sun.  Physical Review B 81, 16 165425 (2010).

13. C. W. J. Beenakker, Physical review letters 97, 6, 067007 (2006).

14. S. G. Cheng, Y. Xing, J. Wang, and Q. F. Sun,  Physical review letters, 103(16), 167003 (2009).

15. Q. F. Sun, J. Wang, and T.H. Lin.  Physical Review B 59, 5, 3831 (1999).

\clearpage

\end{document}